\documentclass[letterpaper, 10 pt, conference]{ieeeconf}  

\IEEEoverridecommandlockouts                              

\usepackage{graphics,graphicx}
\usepackage{array, booktabs, makecell}

\title{\LARGE \bf
Classification of Perceived Human Stress using Physiological Signals
}

\author{Aamir Arsalan$^{1}$, Muhammad Majid$^{1}$, Syed Muhammad Anwar$^{2}$, and Ulas Bagci $^{2}$
\thanks{$^{1}$Aamir Arsalan and $^{1}$Muhammad Majid are with the \textbf{S}ignal, \textbf{I}mage, \textbf{M}ultimedia, \textbf{P}rocessing and \textbf{LE}arning (SIMPLE) Research Group, Department of Computer Engineering, University of Engineering and Technology, Taxila, 47080 Pakistan.
        {\tt\small aamir.arsalan@uettaxila.edu.pk, m.majid@uettaxila.edu.pk}}
\thanks{$^{2}$Syed Muhammad Anwar and Ulas Bagci are with the Center for Research in Computer Vision (CRCV), University of Central Florida, Orlando, Florida, 32816 USA.
        {\tt\small s.anwar@knights.ucf.edu, bagci@ucf.edu}}
}

\def\Fig#1{Fig.~\ref{fig:#1}}

\def\Tab#1{TABLE~\ref{tab:#1}}
\def\Sec#1{Section~\ref{sec:#1}}

\begin{document}

\maketitle
\thispagestyle{empty}
\pagestyle{empty}

\begin{abstract}

In this paper, we present an experimental study for the classification of perceived human stress using non-invasive physiological signals. These include electroencephalography (EEG), galvanic skin response (GSR), and photoplethysmography (PPG). We conducted experiments consisting of steps including data acquisition, feature extraction, and perceived human stress classification. The physiological data of $28$ participants are acquired in an open eye condition for a duration of three minutes. Four different features are extracted in time domain from EEG, GSR and PPG signals and classification is performed using multiple classifiers including support vector machine, the Naive Bayes, and multi-layer perceptron (MLP). The best classification accuracy of $75\%$ is achieved by using MLP classifier. Our experimental results have shown that our proposed scheme outperforms existing perceived stress classification methods, where no stress inducers are used.  

\end{abstract}

\section{INTRODUCTION}

Human stress is a common problem effecting day to day life of a major part of our society. It is also the cause of many chronic diseases. According to a report of American Psychological Association (APA) in 2014, a large population in the United States of America experienced different kinds of physical and psychological symptoms, which were mainly caused by stress \cite{american2015stress}. These included headache, fatigue, lack of clarity in work, and upset stomach. Human stress has been categorized into either acute (instantaneous) or chronic (long-term or perceived). Chronic stress occurs due to events occurring for a relatively longer duration of time in the recent past. It is an established fact that chronic stress acts as a catalyst in the onset of different diseases \cite{cohen2012chronic}. The quality of life is degraded due to these factors causing a reduction in work efficiency of individuals. Various diverse reasons for stress were identified in the APA report such as bad relationship, workplace pressure and economic crises to name a few. Stress for a longer duration of time results in an increased probability of the occurrence of diseases such as heart attack and depression \cite{duman2014neurobiology}. Therefore, it is important to develop a reliable system, which is capable of detecting human stress.

Human stress has been measured in a number of ways (subjective and objective) over the past years. Subjective measures includes questionnaires which are developed by psychologists, such as perceived stress scale (PSS) \cite{cohenperceived} and state-trait anxiety inventory (STAI) \cite{spielberger2017state}. These methods are self reporting measures, which are prone to failure in cases, where an individual answers the questions incorrectly. In these situations, objective methods (physical as well as physiological measures) could be useful. Physical measure for detecting stress could include facial expressions \cite{deschenes2015facial} and eye blink rate \cite{gowrisankaran2012asthenopia}. Whereas, common physiological measures include electroencephalography (EEG) \cite{asif2019human}, heart rate (HR), heart rate variability (HRV) \cite{pereira2017heart}, galvanic skin response (GSR) \cite{panigrahy2017study}, and photoplethysmography (PPG) \cite{chauhan2018real}. In literature, different types of physiological human stress measurement methods have been developed. Most of these methods are for the measurement of acute stress, which is commonly induced by the use of standard stressors such as the stroop color and word test \cite{tulen1989characterization}, trier social stress test (TSST) \cite{kirschbaum1993trier}, and mental arithmetic tasks \cite{sulaiman2012development}. On the other hand, only a few studies are available, which have measured or classified chronic stress in humans \cite{saeed2015psychological,saeed2017quantification,hamid2015brainwaves}. 

In \cite{saeed2015psychological}, a study to establish a relationship between the perceived stress and the brain activity of the participants was presented. The authors concluded that individuals with high perceived stress have an increased level of beta activity in the brain. A regression analysis was presented to predict PSS score of an individual with a confidence level of 94\% using beta activity of the recorded EEG data in closed eye condition \cite{saeed2017quantification} . It was shown that energy spectral density in the right and left hemisphere of the brain of an stressed individual is significantly different from a non-stressed individual \cite{hamid2015brainwaves}. An alpha asymmetry based study to find the relationship between stress and EEG was performed in \cite{sulaiman2011intelligent}. The study reported that individuals with a low level of chronic stress have a dominant left hemisphere, whereas high chronic stress subjects have a dominantly active right hemisphere of the brain. In \cite{hamid2010evaluation}, a correlation based analysis was performed and it was reported that high PSS score individuals have negative alpha and beta activity ratio and vice versa. In \cite{luijcks2015influence}, it was reported that a subject with high PSS score have an increased beta and theta activity in post stimuli phase. In \cite{panigrahy2017study}, a GSR based study for measuring chronic stress of an individual was presented. The GSR data of the participants was recorded in three different states i.e., sitting, standing and sleeping. Human stress was classified to two classes using supervised machine learning algorithms. A PPG based human stress measurement method was presented in \cite{chauhan2018real}. The experimental study induced stress using the paced auditory serial addition test (PASAT). Discrete wavelet transform (DWT) coefficient were extracted from the observed data and adaboost ensemble classifier was used for stress classification.

Most physiological signals based human stress studies available in literature have focused on acute stress, whereas very few experimental methods are available for measuring chronic stress. There are a few studies that have used a single source of physiological signal to establish a relationship between PSS score and the signal used. To the best of our knowledge, no study to classify chronic stress using a combination of EEG, GSR and PPG signal has been conducted so far. In this study, a PSS questionnaire was filled by the participants. They were labeled into stressed and non-stressed groups based on their PSS score. Physiological signals of the participants were acquired after the filling of questionnaires. Four feature groups were extracted from the acquired physiological data and classification was performed using SVM, NB and MLP classifiers. The rest of the paper is organized as follows. \Sec{mes} presents the experimental setup used to classify the stress level. In \Sec{er}, a discussion of the results is presented, followed by conclusion in \Sec{conc}.

\begin{figure}[t]
\begin{center}
\includegraphics[width=85mm]{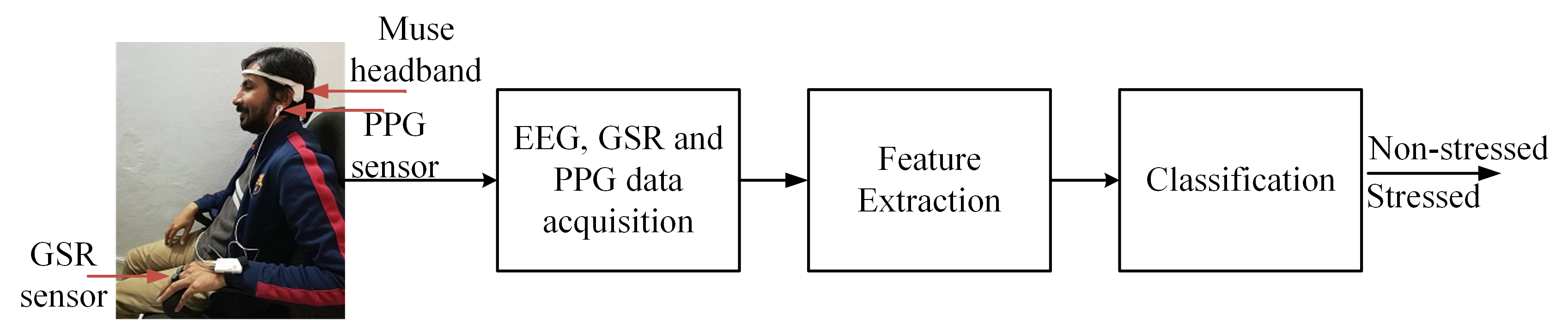}
\caption
{\label{fig:fig1}
A block diagram of the proposed chronic stress classification methodology using physiological signals.}
\end{center}
\end{figure}     

\section{Methodology and Experimental Setup}\label{sec:mes}

Our proposed methodology (\Fig{fig1}) consist of three steps i.e., physiological data acquisition, feature extraction and classification. The details of each step are described in following subsections. 

\subsection{Data Acquisition}

A total of $28$ subjects ($13$ males and $15$ females) with age ranging from $18$ to $40$ years participated in the experiment. The Institution’s Ethical Review Board approved all experimental procedures involving human subjects. None of the participants have reported any mental or physical disabilities. The subjects were initially instructed regarding the experimental procedure, following which the PSS questionnaire was filled by each participant. PSS is a questionnaire developed by psychologists to assess the level of chronic stress faced by an individual. The questionnaire consists of ten questions which can be answered on a numerical scale from $0$ to $4$, where $0$ means the event never occurred and $4$ means that the event occurred very frequently in the last $30$ days. The total PSS score of a subject could range from $0$ to $40$. A score of $0$ means there was no perceived stress and a score of $40$ suggests a high level of perceived stress. 

The EEG, GSR and PPG data for each participant was acquired after filling the questionnaire. EEG data was recorded by using a four channel Interaxon MUSE headband, which is an easy to wear EEG device. The EEG electrodes are located at $TP9$, $AF7$, $AF8$, and $TP10$ positions. The GSR and PPG data was recorded using a Shimmer GSR+ module and a PPG optical pulse clip. The GSR sensor was placed on the index finger and PPG optical pulse clip was placed on the left ear lobe. The sampling rate of EEG headband and GSR+ module was $256$ Hz. The data were acquired for a duration of three minutes with participants sitting in an open eye condition on a relaxed chair in a relatively noise free environment with good lighting condition.

\subsection{Feature Extraction}
A total of four time domain features (kurtosis, entropy, standard deviation to mean absolute ratio, and variance) were extracted from the acquired EEG, GSR and PPG signals. The detail of these features are as follows,  

\noindent \textbf{Kurtosis:}
Kurtosis is a statistical measure used to describe the distribution of data. It is the ratio of the fourth moment and the second moment squared and was calculated as,
\begin{equation} 
K=\sum_{i=1}^{n}n\frac{(X_i-X_{avg})^4}{((X_i-X_{avg})^2)^2},
\end{equation}
where $X_i$ is the $i^{th}$ value of the vector $X$ and $n$ is the number of observations. 

\noindent\textbf{Entropy:}
Entropy is a statistical measure of the randomness of the variable $X$, which was calculated as,
\begin{equation}
H(X)=-\sum_{i=1}^{n}P(x_i)log_2 P(x_i),
\end{equation}
where $P(x_i)$ is the probability mass function of the random variable $X$. 

\noindent\textbf{Standard Deviation to Mean Absolute Ratio:}
Standard deviation to mean absolute ratio for the data of each EEG channel as well as GSR and PPG data was calculated as,
\begin{equation}
Ratio=\frac{\sigma_i}{\mu_i},
\end{equation}   
where $\sigma_i$ is the standard deviation and $\mu_i$ is the mean of EEG, GSR or PPG data and the value of $i$ can be $TP9$, $AF7$, $AF8$ or $TP10$ for EEG data. 

\noindent\textbf{Variance:}
Variance of the data for each channel of EEG, GSR and PPG was calculated as,
\begin{equation}
\sigma^2 = \frac{\Sigma(X_i-\mu_i)^2}{n}
\end{equation} 
where $\sigma^2$ is the variance, $X_i$ is the $i^{th}$ value of the vector, $\mu_i$ is the mean of the vector $i$. A total of six feature values were calculated. This included one value for each of the four EEG channels (AF7, AF8, TP9, and TP10). A feature vector of length $1 \times 24$ was obtained with one value for both GSR and PPG data calculated for each feature group (kurtosis, entropy, devation to mean absolute ratio, and variance). Feature selection was applied to the extracted feature using the wrapper method \cite{mustaqeem2017wrapper}. 
The boxplot of the feature values from four feature groups for non-stressed and stresses classes are shown in Fig. It is evident that these four feature groups are significantly different for both the stressed and non-stressed groups, therefore used in this study.  

\begin{figure*}
    \centering
    \includegraphics[width = 140mm]{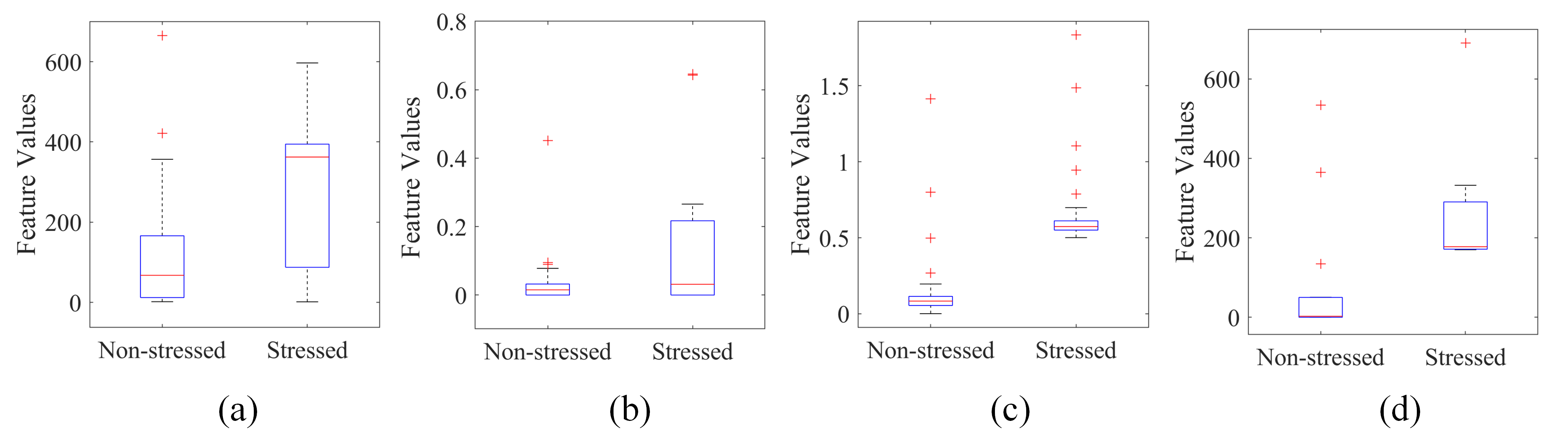}
    \caption{Boxplot for features from the selected feature groups for both stressed and non-stressed subjects.}
    \label{fig:my_label}
\end{figure*}

\subsection{Classification}
Three classifiers were used to perform a binary classification of chronic stress using the EEG, GSR and PPG signals as described in the following sub-sections.

\noindent\textbf{Support Vector Machine (SVM):}
In SVM, the data points are placed in an n-dimensional space and the algorithm finds a hyper plane that separate the data points into two classes. The SVM is an iterative algorithm, which only terminates when a specified minimum error value is achieved. The points near the hyper plane are termed as support vectors.     

\noindent\textbf{Naive Bayes (NB):}
The Naive Bayes is a classification algorithm based on Bayes theorem with an assumption of independence among features. The algorithm works for both binary and multi-class classification problems and for continuous and categorical data. Despite its simplicity, it often outperforms other more sophisticated classifiers. 

\noindent\textbf{Multilayer Perceptron (MLP):}
MLP (artificial neural network (ANN)) can be viewed as a logistic regression based classifier, which transforms the input using a non-linear transformation. The input data becomes separable as a result of this transformation. 
The model is trained on a set of input-output pairs to develop a correlation between them by adjusting the weights in order to minimize the error function. 

\section{Experimental Results}\label{sec:er}
The perceived stress was classified using a combination of EEG, GSR, and PPG signals. Muse monitor application was used for EEG data acquisition from Muse headband. The GSR and PPG data was recorded using Shimmer Capture (v0.6). The subjects with a PSS score ranging from $0-20$ were labeled as non-stressed, whereas the participants with a score greater than $20$ were labeled as stressed. This labeled data was used for the purpose of classifier training. Four groups of time domain features were extracted from the recorded physiological signals and stress was classified into two levels using three different classifiers. We observed that the wrapper based feature selection method reduced the feature vector (24 feature values) to 16 values. The classification experiments were performed on Weka 3.8.3. 

\begin{figure}[!t]
\begin{center}
\includegraphics[width=85mm]{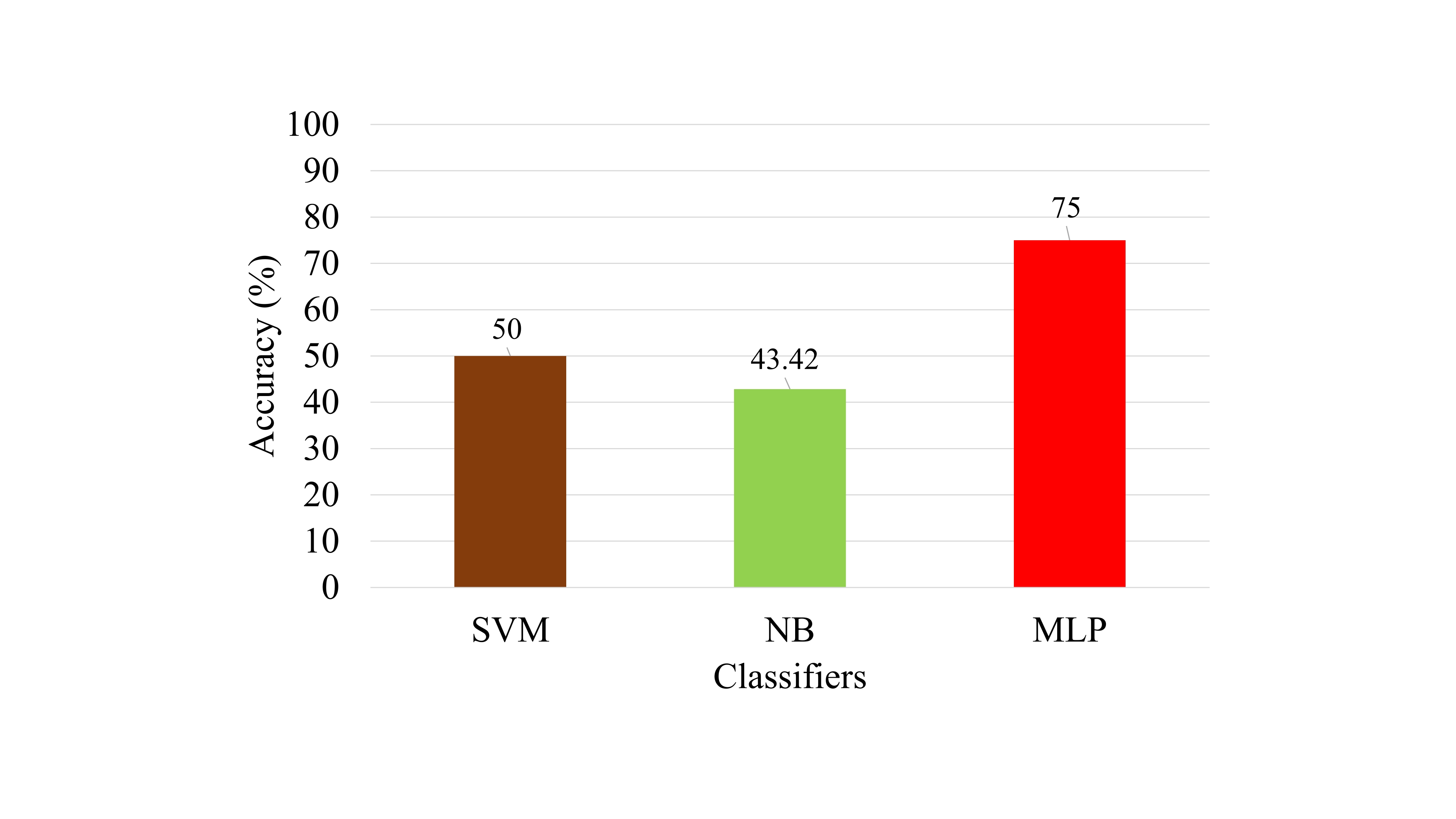}
\caption
{\label{fig:fig3}
Accuracy of three different classifiers used to classify chronic stress using physiological signal based features.}
\end{center}
\end{figure}

\begin{figure}[!t]
\begin{center}
\includegraphics[width=85mm]{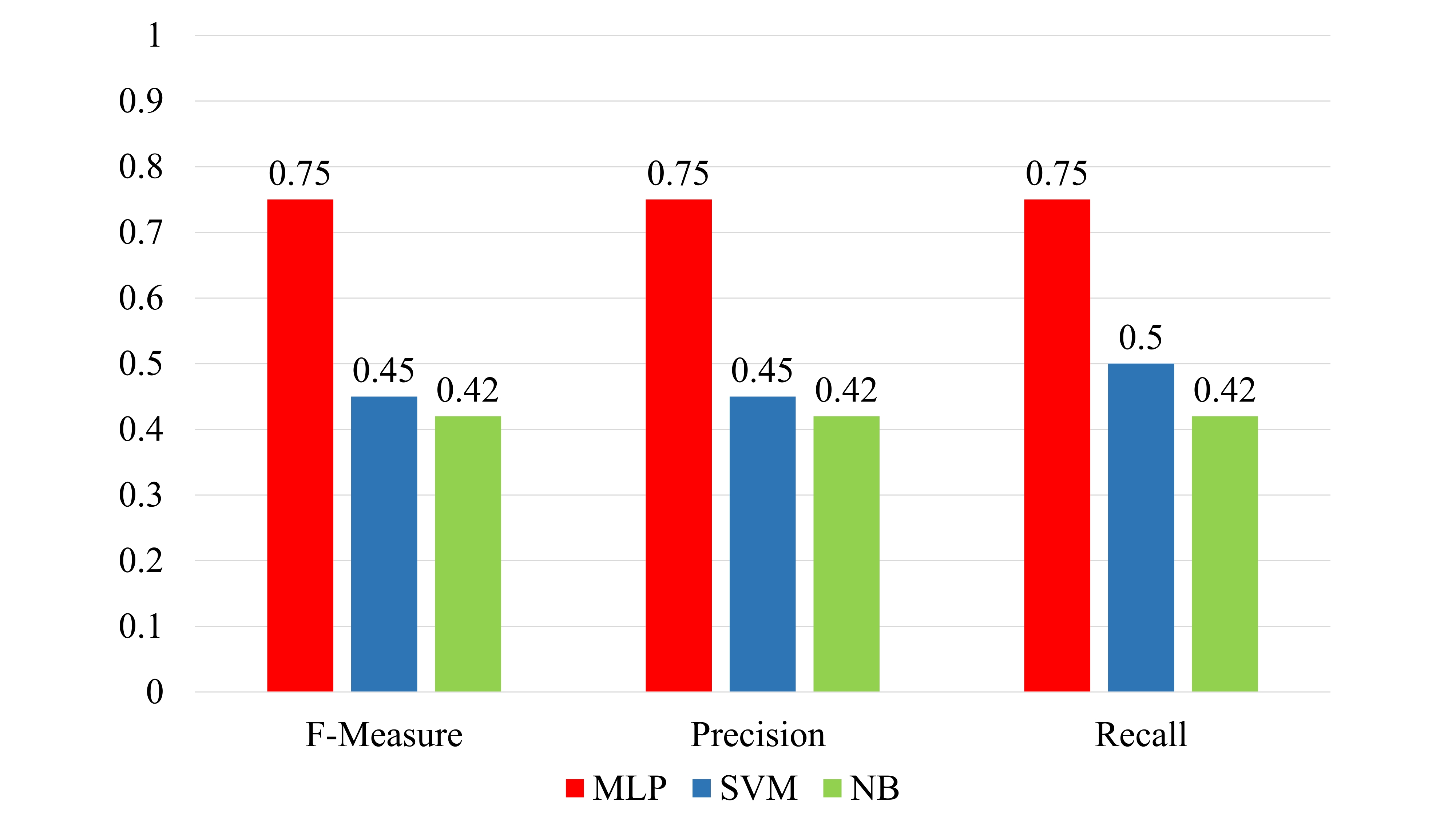}
\caption
{\label{fig:fig4}
A comparison of different classifiers in terms of Precision, Recall and F-measure for chronic stress classification using physiological signals.}
\end{center}
\end{figure}
A $10$-fold cross-validation was applied to evaluate the classifier performance in terms of accuracy, precision, recall and F-measure. A three layer network was used for MLP and for SVM polynomial kernel was used. \Fig{fig3} shows the average accuracy of different classifiers used to classify stress. It is evident that MLP classifier gave the highest accuracy of $75\%$, whereas the lowest accuracy of $42.85\%$ was achieved by NB. The SVM classifier achieved an accuracy of $50\%$. A performance comparison in terms of precision, recall, and F-measure is presented in \Fig{fig4}. We observe that MLP classifier was better in terms of F-measure, precision and recall as compared to both SVM and NB in stress classification. It can be seen that MLP is able to correctly classify $12$ out of $16$ stressed users and $9$ out of $12$ non-stressed users (\Tab{tab1}).

\begin{table}
\caption{Confusion Matrix for MLP classifier.}
\label{tab:tab1}
\centering
\scalebox{1}{
\begin{tabular}{|c|c|c|c|}
\hline
 {Labeled} / {Classified}&  \thead{Stressed} &  \thead{Non-stressed} & \thead{Individual accuracy} \\ \hline
  \thead{Stressed} & 12 & 4 & 75\%\\ \hline
  \thead{Non-stressed} & 3 & 9 & 75\%\\ \hline

\end{tabular}
}
\end{table}

\begin{table}
\caption{Performance comparison of the proposed scheme with state-of-the-art methods.}

\label{tab:tab2}
\centering
\scalebox{0.7}{
\begin{tabular}{|c|c|c|c|c|c|}
\hline
\thead{Method} &  \thead{Number of \\ Participants} &  \thead{Modality \\ Used} & Accuracy (Classes) & Classifier & \thead{Stress \\ Type}\\ \hline
\thead{Sanay et. al. \cite{saeed2017quantification}} & 28 & EEG & 71.40\% (2) & \thead{Naive \\ Bayes} & Perceived  \\
\hline
\thead{Proposed } & 28 & \thead{GSR} & 35.71\% (2) & MLP & Perceived \\
\thead{Proposed } & 28 & \thead{PPG} & 64.28\% (2) & MLP & Perceived \\
\thead{Proposed } & 28 & \thead{EEG, GSR, \\ PPG} & 75\% (2) & MLP & Perceived \\
\hline
\end{tabular}
}
\end{table}  

\Tab{tab2} presents the performance comparison of the proposed algorithm with recent methods available in literature for the classification of human perceived stress. We observed that only a few studies are available for the quantification of perceived stress. The proposed scheme was compared in terms of number of participants in the experiment, modality, and accuracy. Although, the number of participants in \cite{saeed2017quantification,saeed2015psychological} and the proposed scheme is similar ($28$) but in earlier studies only EEG was used for the classification of perceived stress. Whereas, in our proposed scheme a combination of EEG, GSR and PPG signal was used. Perceived stress classification accuracy, for only GSR and only PPG features, was $35.71\%$ and $64.28\%$, respectively. A classification accuracy of $71.4\%$ was achieved when only EEG was used for classification. Whereas, the combination of EEG, GSR and PPG in our proposed method increased the classification accuracy to $75\%$. These results would contribute in the development of a wearable solution for observing long term stress in real life situations. This could ultimately lead towards a brain computer interface based system for effective stress management in daily life.   

\section{Conclusion}\label{sec:conc}
In this paper, an experimental study to classify perceived human stress using a combination of EEG, GSR and PPG signals was presented. Four groups of time domain features namely, entropy, kurtosis, standard deviation to mean absolute ratio and variance were extracted from the EEG, GSR and PPG data recorded in an open eye condition. Three different machine learning algorithms were used to classify the perceived stress and it is concluded that MLP classifier gives the best classification performance as compared to other classifiers with an accuracy of $75\%$, which was higher than the accuracy achieved by only using the EEG data. In future, we intend to use frequency domain features of EEG along with PPG and GSR data to increase the classification accuracy of the algorithm.     

\bibliographystyle{IEEEtran}
\bibliography{mybibfile}

\begin{thebibliography}{10}
\providecommand{\url}[1]{#1}
\csname url@rmstyle\endcsname
\providecommand{\newblock}{\relax}
\providecommand{\bibinfo}[2]{#2}
\providecommand\BIBentrySTDinterwordspacing{\spaceskip=0pt\relax}
\providecommand\BIBentryALTinterwordstretchfactor{4}
\providecommand\BIBentryALTinterwordspacing{\spaceskip=\fontdimen2\font plus
\BIBentryALTinterwordstretchfactor\fontdimen3\font minus
  \fontdimen4\font\relax}
\providecommand\BIBforeignlanguage[2]{{%
\expandafter\ifx\csname l@#1\endcsname\relax
\typeout{** WARNING: IEEEtran.bst: No hyphenation pattern has been}%
\typeout{** loaded for the language `#1'. Using the pattern for}%
\typeout{** the default language instead.}%
\else
\language=\csname l@#1\endcsname
\fi
#2}}

\bibitem{american2015stress}
A.~P. Association \emph{et~al.}, ``Stress in america: Paying with our health,''
  \emph{Retrieved May}, vol.~29, p. 2015, 2015.

\bibitem{cohen2012chronic}
S.~Cohen, D.~Janicki-Deverts, W.~J. Doyle, G.~E. Miller, E.~Frank, B.~S. Rabin,
  and R.~B. Turner, ``Chronic stress, glucocorticoid receptor resistance,
  inflammation, and disease risk,'' \emph{Proceedings of the National Academy
  of Sciences}, vol. 109, no.~16, pp. 5995--5999, 2012.

\bibitem{duman2014neurobiology}
R.~S. Duman, ``Neurobiology of stress, depression, and rapid acting
  antidepressants: remodeling synaptic connections,'' \emph{Depression and
  anxiety}, vol.~31, no.~4, pp. 291--296, 2014.

\bibitem{cohenperceived}
S.~Cohen, T.~Kamarck, and R.~Mermelstein, ``Perceived stress scale. measuring
  stress: A guide for health and social scientists. 1994.''

\bibitem{spielberger2017state}
C.~D. Spielberger, F.~Gonzalez-Reigosa, A.~Martinez-Urrutia, L.~F. Natalicio,
  and D.~S. Natalicio, ``The state-trait anxiety inventory,'' \emph{Revista
  Interamericana de Psicologia/Interamerican Journal of Psychology}, vol.~5,
  no. 3 \& 4, 2017.

\bibitem{deschenes2015facial}
A.~Desch{\^e}nes, H.~Forget, C.~Daudelin-Peltier, D.~Fiset, and C.~Blais,
  ``Facial expression recognition impairment following acute social stress,''
  \emph{Journal of vision}, vol.~15, no.~12, pp. 1383--1383, 2015.

\bibitem{gowrisankaran2012asthenopia}
S.~Gowrisankaran, N.~K. Nahar, J.~R. Hayes, and J.~E. Sheedy, ``Asthenopia and
  blink rate under visual and cognitive loads,'' \emph{Optometry and Vision
  Science}, vol.~89, no.~1, pp. 97--104, 2012.

\bibitem{asif2019human}
A.~Asif, M.~Majid, and S.~M. Anwar, ``Human stress classification using eeg
  signals in response to music tracks,'' \emph{Computers in biology and
  medicine}, vol. 107, pp. 182--196, 2019.

\bibitem{pereira2017heart}
T.~Pereira, P.~R. Almeida, J.~P. Cunha, and A.~Aguiar, ``Heart rate variability
  metrics for fine-grained stress level assessment,'' \emph{Computer methods
  and programs in biomedicine}, vol. 148, pp. 71--80, 2017.

\bibitem{panigrahy2017study}
S.~K. Panigrahy, S.~K. Jena, and A.~K. Turuk, ``Study and analysis of human
  stress detection using galvanic skin response (gsr) sensor inwired and
  wireless environments,'' \emph{Research Journal of Pharmacy and Technology},
  vol.~10, no.~2, pp. 545--550, 2017.

\bibitem{chauhan2018real}
U.~Chauhan, N.~Reithinger, and J.~R. Mackey, ``Real-time stress assessment
  through ppg sensor for vr biofeedback,'' in \emph{Proceedings of the
  International Conference on Multimodal Interaction: Adjunct}.\hskip 1em plus
  0.5em minus 0.4em\relax ACM, 2018, p.~5.

\bibitem{tulen1989characterization}
J.~Tulen, P.~Moleman, H.~Van~Steenis, and F.~Boomsma, ``Characterization of
  stress reactions to the stroop color word test,'' \emph{Pharmacology
  Biochemistry and Behavior}, vol.~32, no.~1, pp. 9--15, 1989.

\bibitem{kirschbaum1993trier}
C.~Kirschbaum, K.-M. Pirke, and D.~H. Hellhammer, ``The ‘trier social stress
  test’--a tool for investigating psychobiological stress responses in a
  laboratory setting,'' \emph{Neuropsychobiology}, vol.~28, no. 1-2, pp.
  76--81, 1993.

\bibitem{sulaiman2012development}
N.~Sulaiman, M.~N. Taib, S.~Lias, Z.~H. Murat, S.~A.~M. Aris, M.~Mustafa, and
  N.~A. Rashid, ``Development of eeg-based stress index,'' in \emph{Biomedical
  Engineering (ICoBE), 2012 International Conference on}.\hskip 1em plus 0.5em
  minus 0.4em\relax IEEE, 2012, pp. 461--466.

\bibitem{saeed2015psychological}
S.~M.~U. Saeed, S.~M. Anwar, M.~Majid, and A.~M. Bhatti, ``Psychological stress
  measurement using low cost single channel eeg headset,'' in \emph{Signal
  Processing and Information Technology (ISSPIT), 2015 IEEE International
  Symposium on}.\hskip 1em plus 0.5em minus 0.4em\relax IEEE, 2015, pp.
  581--585.

\bibitem{saeed2017quantification}
S.~M.~U. SAEED, S.~M. ANWAR, and M.~Majid, ``Quantification of human stress
  using commercially available single channel eeg headset,'' \emph{IEICE
  Transactions on Information and Systems}, vol. 100, no.~9, pp. 2241--2244,
  2017.

\bibitem{hamid2015brainwaves}
N.~H.~A. Hamid, N.~Sulaiman, Z.~H. Murat, and M.~N. Taib, ``Brainwaves stress
  pattern based on perceived stress scale test,'' in \emph{Control and System
  Graduate Research Colloquium (ICSGRC), 2015 IEEE 6th}.\hskip 1em plus 0.5em
  minus 0.4em\relax IEEE, 2015, pp. 135--140.

\bibitem{sulaiman2011intelligent}
N.~Sulaiman, M.~N. Taib, S.~Lias, Z.~H. Murat, S.~A.~M. Aris, M.~Mustafa, N.~A.
  Rashid, and N.~H.~A. Hamid, ``Intelligent system for assessing human stress
  using eeg signals and psychoanalysis tests,'' in \emph{Computational
  Intelligence, Communication Systems and Networks (CICSyN), 2011 Third
  International Conference on}.\hskip 1em plus 0.5em minus 0.4em\relax IEEE,
  2011, pp. 363--367.

\bibitem{hamid2010evaluation}
N.~H.~A. Hamid, N.~Sulaiman, S.~A.~M. Aris, Z.~H. Murat, and M.~N. Taib,
  ``Evaluation of human stress using eeg power spectrum,'' in \emph{Signal
  Processing and Its Applications (CSPA), 2010 6th International Colloquium
  on}.\hskip 1em plus 0.5em minus 0.4em\relax IEEE, 2010, pp. 1--4.

\bibitem{luijcks2015influence}
R.~Luijcks, C.~J. Vossen, H.~J. Hermens, J.~van Os, and R.~Lousberg, ``The
  influence of perceived stress on cortical reactivity: a proof-of-principle
  study,'' \emph{PloS one}, vol.~10, no.~6, p. e0129220, 2015.

\bibitem{mustaqeem2017wrapper}
A.~Mustaqeem, S.~M. Anwar, M.~Majid, and A.~R. Khan, ``Wrapper method for
  feature selection to classify cardiac arrhythmia,'' in \emph{2017 39th Annual
  International Conference of the IEEE Engineering in Medicine and Biology
  Society (EMBC)}.\hskip 1em plus 0.5em minus 0.4em\relax IEEE, 2017, pp.
  3656--3659.

\end{thebibliography}

\end{document}